\documentclass[aps,prl,twocolumn,superscriptaddress,showpacs,floatfix]{revtex4-1}

\usepackage{graphicx}
\usepackage{textcomp}
\usepackage{siunitx}
\usepackage{bm}
\usepackage{color}

\newcommand{\bnabla}{\mbox{\boldmath $\nabla$}}
\newcommand{\bdiv}{\bnabla\cdot}
\newcommand{\bcurl}{\bnabla\times}
\newcommand{\dd}{\mathrm{d}}

\newcommand{\ve}[1]{\bf #1}
\newcommand{\veh}[1]{\bm{\mathbf{\hat{#1}}}}

\begin{document}

\title{Chiral structures and defects of lyotropic chromonic liquid crystals induced by saddle-splay elasticity}

\author{Zoey S.\ Davidson}
\thanks{J.J., L.K., and Z.S.D. contributed equally to this work.}
\author{Louis Kang}
\thanks{J.J., L.K., and Z.S.D. contributed equally to this work.}
\affiliation{Department of Physics \& Astronomy, University of Pennsylvania, Philadelphia, PA 19104, USA}
\author{Joonwoo Jeong}
\email{jjeong@unist.ac.kr}
\affiliation{School of Natural Science, Department of Physics, Ulsan National Institute of Science and Techonlogy (UNIST), Ulsan 689-798, Republic of Korea}
\author{Tim Still}
\affiliation{Department of Physics \& Astronomy, University of Pennsylvania, Philadelphia, PA 19104, USA}
\author{Peter J.\ Collings}
\affiliation{Department of Physics \& Astronomy, Swarthmore College, Swarthmore, PA 19081, USA}
\affiliation{Department of Physics \& Astronomy, University of Pennsylvania, Philadelphia, PA 19104, USA}
\author{Tom C.\ Lubensky}
\author{A.~G.\ Yodh}
\affiliation{Department of Physics \& Astronomy, University of Pennsylvania, Philadelphia, PA 19104, USA}

\date{\today}

\begin{abstract}
An experimental and theoretical study of lyotropic chromonic liquid crystals (LCLCs) confined in cylinders with degenerate planar boundary conditions elucidates LCLC director configurations. When the Frank saddle-splay modulus is more than twice the twist modulus, the ground state adopts an inhomogeneous escaped-twisted configuration. Analysis of the configuration yields a large saddle-splay modulus, which violates Ericksen inequalities but not thermodynamic stability. Lastly, we observe point defects between opposite-handed domains and explain a preference for point defects over domain walls.
\end{abstract}

\pacs{61.30.Jf, 61.30.Pq, 62.20.de, 68.37.-d}

\maketitle

The elastic properties of nematic liquid crystals (LCs) are crucial for liquid crystal display applications \cite{schadt_voltagedependent_1971,yeh_optics_2010}, and they continue to give rise to unanticipated fundamental phenomena \cite{jeong_chiral_2014,jeong_chiral_2015,skarabot_hierarchical_2008,rao_low_2009,gibaud_reconfigurable_2012,jones_bistable_2012,li_liquid_2012}. Three of the bulk nematic LC deformation modes, splay, twist and bend, are well known and have associated elastic moduli $K_{1}$, $K_{2}$ and $K_{3}$, respectively. These moduli have been intensely studied because they are easy to visualize, and because it is possible to independently excite the modes via clever usage of sample geometry \cite{hakemi_determination_1983,giavazzi_viscoelasticity_2014,zhou_elasticity_2014}, LC boundary conditions \cite{sparavigna_periodic_1994,lavrentovich_geometrical_1992}, and external fields \cite{freedericksz_forces_1933,zhou_elasticity_2012}. As a result, these moduli have been measured for a variety of thermotropic and lyotropic LCs \cite{karat_elastic_1976,karat_elasticity_1977,madhusudana_elasticity_1982,bogi_elastic_2001,zhou_elasticity_2012, zhou_elasticity_2014}. By contrast, a much less studied fourth independent mode \cite{nehring_elastic_1971,crawford_saddle-splay_1995, schmidt_normal-distortion-mode_1990} of elastic deformation in nematic LCs can exist; it is called saddle-splay. Saddle-splay is hard to visualize and to independently excite~\cite{schmidt_normal-distortion-mode_1990,palffy-muhoray_comment_1990}. Moreover, the energy of this deformation class can be integrated to the boundary, so that the mode does not appear in the Euler-Lagrange equations, and with fixed boundary conditions, the saddle-splay energy will have no effect on the LC director configuration. Even with free boundary conditions, the saddle-splay energy will not affect the bulk LC configuration unless the principal curvatures of the surface are different, i.e., saddle-splay effects are not expected for spherical or flat surfaces. Thus, although much progress in understanding saddle-splay has been made \cite{gennes_physics_1995,alexander_disclination_2012}, especially with thermotropic nematic LCs, unambiguous determination of saddle-splay energy effects on liquid crystal configurations and measurement of  the saddle-splay elastic modulus, $K_{24}$, remain difficult \cite{joshi_measuring_2014}.

While the bulk elastic constants described above strongly influence LC director configurations, LC boundary conditions at material interfaces also influence bulk structure. Indeed, considerable effort has gone into development of surface preparation techniques to produce particular bulk director configurations \cite{schadt_surface-induced_1992, nazarenko_surface_2010, zimmermann_self-organized_2015, lee_alignment_2001, willman_liquid_2014, sparavigna_periodic_1994, mcginn_planar_2013}. The saddle-splay term integrates to the boundary and effectively imposes boundary conditions at free surfaces favoring director alignment along the direction of highest surface curvature for positive $K_{24}$ \cite{koning_saddle-splay_2014} and outwardly pointing surface normals. For this effect to be present, the director cannot be held perpendicular to the surface, as was the case in our prior work \cite{jeong_chiral_2015}. The potential role of saddle-splay effects in determining bulk director configurations by spontaneous symmetry breaking has been appreciated \cite{pairam_stable_2013,crawford_surface_1992,hough_helical_2009,sparavigna_periodic_1994} but has been difficult to fully characterize; generally, molecular surface forces can impose preferred boundary conditions that are hard to disentangle from effects due to $K_{24}$ \cite{ondris-crawford_curvature-induced_1993,melzer_optical_1977}. As a result, the measurements of $K_{24}$ to date have wide confidence intervals  \cite{allender_determination_1991,pairam_stable_2013,polak_optical_1994} and even vary in sign \cite{polak_optical_1994}. Finally, additional factors that have complicated assignment of saddle-splay effects are the so-called Ericksen inequalities \cite{ericksen_inequalities_1966} that require $0 < K_{24} <2 K_2$ and $K_{24} < 2 K_1$. These inequalities were derived assuming spatially uniform gradients of the director. They do not, however, apply in geometries such as ours in which gradients are not uniform.  

In this contribution, we investigate director configurations of the nematic lyotropic chromonic liquid crystal (LCLC) Sunset Yellow (SSY) confined within cylindrical glass capillaries with degenerate planar boundary conditions as initially reported by Refs. \cite{davidson2014lyotropic, davidson2015planar, chang2015ground}. Our study employs a combination of polarized optical microscopy, measurements of director-field thermal fluctuations, and Frank-free-energy calculations to rationalize the observed structures. Importantly, we show that a large $K_{24}$ leads to an escaped-twist (ET) ground state, which exhibits a classic double-twist configuration.  Note, chiral symmetry breaking in the ET configuration is fundamentally different from symmetry breaking in other LCLC systems with uniform principal curvatures \cite{jeong_chiral_2014}, or with homeotropic boundary conditions \cite{jeong_chiral_2015}. In the previous work, spontaneous twist deformation arises because $K_{2}$ is much smaller than $K_{1}$ and $K_{3}$; $K_{24}$ played no role in the energetics. In the present work, comparison of theory and experiment enables us to measure $K_{24}$ for the first time in a LCLC; we find a value of $K_{24}/K_{2} = 55.0$, which strongly violates the Ericksen inequalities. Finally, we observe and characterize chiral hegdehog point defects separating chiral domains of opposite handedness. Interestingly, the presence of point defects rather than smooth domain walls also provides precise quantitative information about $K_{24}$ that is consistent with our other conclusions.

\begin{figure}[b]
 \includegraphics[width=1.0\linewidth]{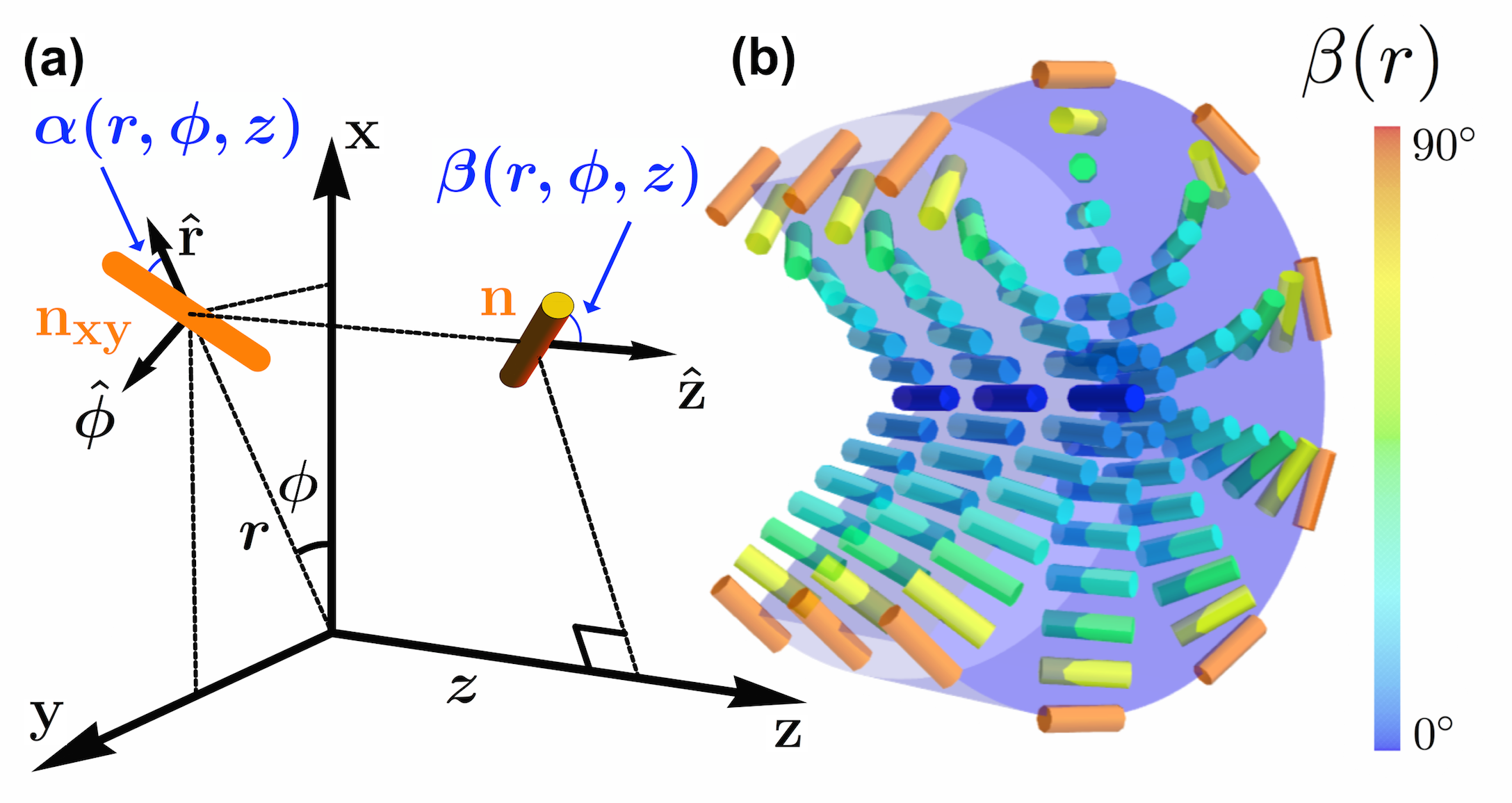}
 \caption{\label{Fig1} (Color online) {\bf (a)}  Coordinate system used for director configuration and defect energy calculations. The director $\ve n$ is described by the angle $\alpha$ between the director projection $\ve n_{xy}$ and $\veh r$, and by the angle $\beta$ between $\ve n$ and the capillary axis, which is parallel to $\veh z$. {\bf (b)} 3D cutaway view of the capillary and the ground state director field using $K_2/K_3 = 0.1$, and $K_{24}/K_{3} = 4.6$, which approximates the Frank moduli of 30 \% wt./wt. SSY at $25^{\circ}$C. Notice the large twist angle at the capillary surface close to $90^{\circ}$. $\alpha$ is independent of position and $\beta$ depends on only the radial coordinate $r$. This configuration has right-handed chirality.}
 \end{figure}

Before discussing the experimental results, we formulate the theoretical problem. We assume the achiral nematic LCLC is described by a Frank free energy, i.e.,  
\begin{widetext}
\begin{equation}
	F = \int \dd^3 {\ve x} \left[ \frac{1}{2} K_{1} \left( \ve n \bdiv \ve n  \right) ^{2} + \frac{1}{2} K_{2} \left( \ve n \cdot \bcurl \ve n \right)^2 + \frac{1}{2} K_{3} \left( \ve n \times \bcurl \ve n \right)^2 - \frac{1}{2} K_{24} \bdiv \left(\ve n \times \bcurl \ve n + \ve n \bdiv \ve n\right) \right],
	\label{eqn:fgen}
\end{equation}
\end{widetext} 
where $\ve n$ is the nematic director. Equation (\ref{eqn:fgen}) explicitly includes the saddle-splay term with modulus $K_{24}$, which can in principle be mimicked by a surface anchoring term that is coupled to surface curvature; thus we consider a saddle-splay term that combines the two effects \cite{PhysRevE.73.051706,NoteX}. A Rapini-Papoular type surface anchoring term with in-plane anisotropy \cite{ondris-crawford_curvature-induced_1993,crawford_saddle-splay_1995} is excluded and discussed later in the text. The LC is contained inside a capillary of radius $R$ and cylindrical coordinates are used to parameterize its director field, $\ve n$, with $\hat{{\bf z}}$ along the capillary axis (see Fig.~\ref{Fig1}), i.e., 
\begin{equation}
	\ve n = \cos\alpha\sin\beta\,\veh r + \sin\alpha\sin\beta\,\veh \phi + \cos\beta\,\veh z.
	\label{eqn:n}
\end{equation}
To determine the configuration of the ground state, we assume the director depends only on $r$ and minimize the Frank free energy with respect to $\alpha(r)$ and $\beta(r)$. Degenerate planar anchoring conditions at the capillary surface prevent the director from having an $\hat{{\bf r}}$-component, so $\alpha(r=R)=\pi/2$. Cylindrical symmetry sets $\beta(r=0)=0$. Both $\alpha(r=0)$ and $\beta(r=R)$ are free to vary, but stationarity of the free energy provides the boundary conditions: $\partial_{r}\alpha(r=0)=0 $ and $R\partial_{r}\beta(r=R)=(\frac{K_{24}}{2K_{2}}- \frac{1}{2}) \sin 2\beta(r=R)$. 

\begin{figure}[b]
 \includegraphics[width=1.0\linewidth]{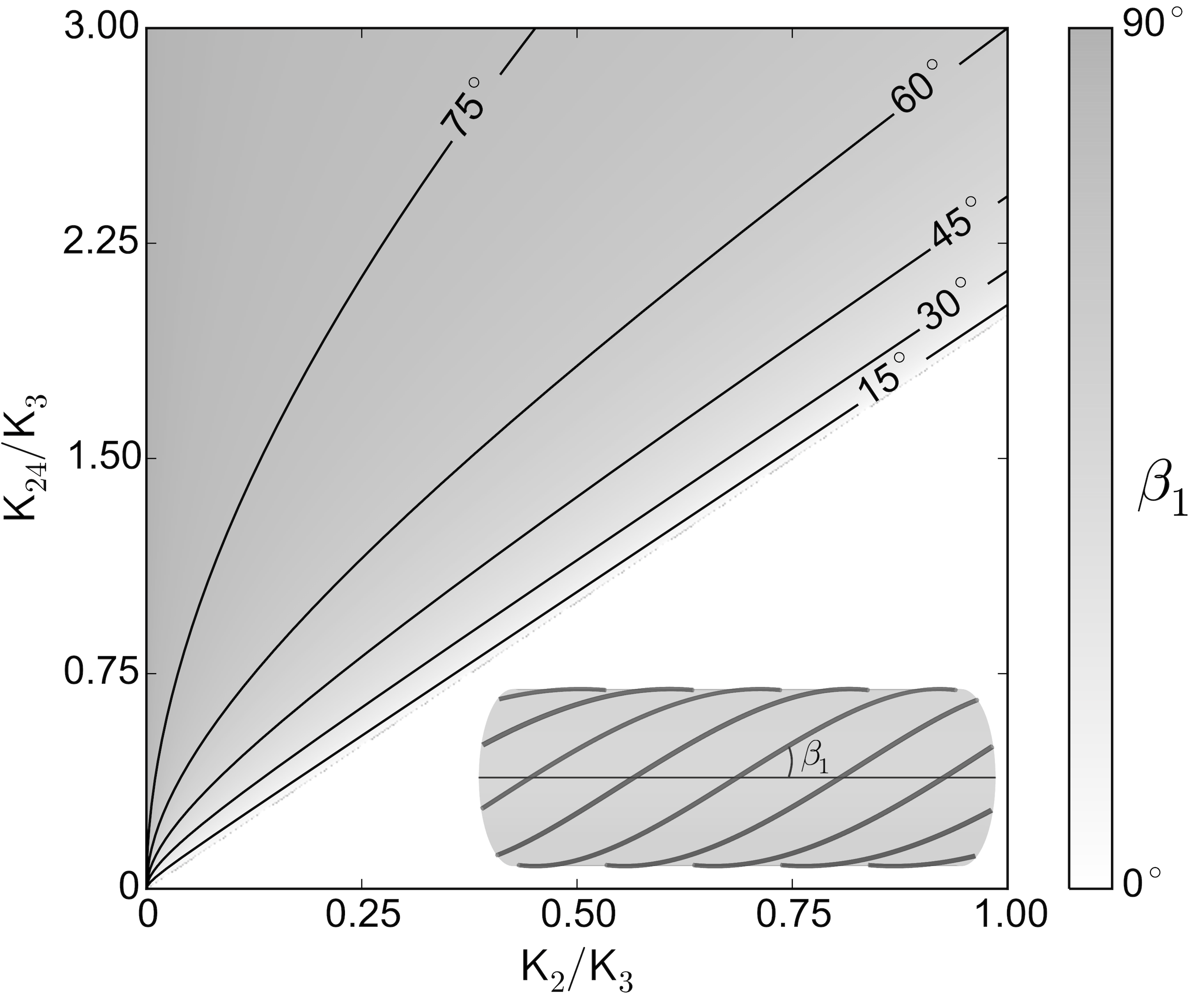}
 \caption{\label{Fig2}Phase diagram of $\beta_{1}$, the angle between the nematic director and capillary axis at the capillary surface, as a function of elastic moduli ratios of saddle-splay $(K_{24})$ to twist $(K_{2})$ and bend $(K_{3})$. Inset: an example capillary with streamlines indicating a surface director field at angle $\beta_{1}$ with left-handed chirality.}
 \end{figure}

With these boundary conditions, the Euler-Lagrange equations of the Frank free energy give \cite{ondris-crawford_curvature-induced_1993}

\begin{eqnarray}
\alpha(r) &= &\frac{\pi}{2};
\\
\beta(r) &= & \arctan\frac{2\sqrt{K_2 K_{24}(K_{24}-2K_2)} r/R}{\sqrt{K_3}[K_{24}-(K_{24}-2K_2)r^2/R^2]}.
\label{eqn:betas}
\end{eqnarray}
This ET solution exists for $K_{24} > 2K_2$ and has right-handed chirality, i.e., the director streamlines form right-handed helices. A mirror-image solution $\beta(r) \rightarrow \pi - \beta(r)$ exists with the same energy. Notice that the radial position $r$ is scaled by the cylinder radius $R$ and that $K_1$ does not appear because this configuration has no splay. If $K_{24} < 2K_{2}$, then only the trivial $\beta(r)=0$ solution exists, which corresponds to the simple parallel-axial configuration \cite{crawford_surface_1992}. As $K_{24}$ surpasses $2K_{2} \equiv K_{c}$, which is exactly the upper bound found by Ericksen, the system spontaneously breaks chiral symmetry, and an ET configuration of one handedness grows continuously from the trivial solution. $\beta_{1}=\beta(r=R)$ is plotted in Fig.~\ref{Fig2}. Prior work with thermotopic LCs has found this ET configuration when an azimuthal anchoring condition dominates the behavior of $\beta_1$ at the capillary surface through a chemical or mechanical treatment of the surface \cite{ondris-crawford_curvature-induced_1993,melzer_optical_1977,cladis_non-singular_1972}. $\beta(r)$ (Eq.~\ref{eqn:betas}) can only be approximated by a linear twist model  \cite{koning_saddle-splay_2014} for certain ratios of elastic constants. For LCs whose elastic moduli do not satisfy these ratios, such as SSY, polarized optical microscopy textures are strongly affected by the nonlinear behavior of $\beta(r)$ \cite{NoteX}.

The normalized free energy of the ET configuration is readily calculated to be
\begin{eqnarray}
\frac{F}{\pi L} = &-&(K_{24}-2K_2) \\
&+& \frac{\sqrt{K_2}K_3}{\sqrt{K_3-K_2}} \arctan\frac{\sqrt{K_3-K_2}(K_{24}-2K_2)}{\sqrt{K_2}(K_3+K_{24}-2K_2)},
\nonumber
\label{eqn:fk24}
\end{eqnarray}
where $L$ is the length of the capillary. Notice that as $K_{24}$ increases beyond $2K_{2}$, the free energy decreases continuously from $0$, thereby confirming that the ET configuration as a ground state is preferred over the uniform configuration whenever it can exist; $K_{24}  = 2K_{2}$ marks a second-order phase transition line. The key to this energetic stabilization is the saddle-splay term:

\begin{equation}
	\frac{F_{24}}{\pi L} = - K_{24}\sin^{2}\beta_{1}.
\label{eqn:ssF24}
\end{equation}
As noted by Ref.\ \cite{koning_saddle-splay_2014}, who use an opposite surface normal convention, $F_{24}$ couples the nematic director to the surface curvature tensor and favors alignment in the direction of highest curvature for $K_{24} > 0$. In our case, this is the azimuthal direction along the circumference of the capillary. Thus, the saddle-splay free energy stabilizes the ET configuration despite introducing bulk director distortion. We also have verified that both the ET and the deformation-free solutions are stable whenever they are preferred ($K_{24} > 2K_2$ and $K_{24} < 2K_2$, respectively); that is, their stability matrices have positive eigenvalues (see Supplemental Information) \cite{NoteX}.

Our experimental investigations used nematic SSY, a LCLC with relatively low twist modulus $K_2/K_3\approx0.1$  \cite{zhou_elasticity_2012}. Briefly, five SSY samples were loaded into five different capillary tubes with diameters 100~\textmu m $\pm 10\%$, from VitroCom (CV1017-100). The sealed samples were illuminated between cross-polarizers by 10nm-bandpass-filtered 660 nm LED light at high (160x) magnification, enabling small depth of field and high spatial resolution imaging. Images were captured by a Uniq UP680-CL video camera, and a piezo-objective positioner was moved to image focal planes within the samples in 1\ \textmu m intervals.

The capillaries without surface treatment were loaded with SSY and sealed to prevent evaporation. A critical experimental question for any saddle-splay study concerns possible structure on the cylinder interfaces that could induce a preferred anchoring direction. To this end, we examined the inner capillary surfaces using atomic force microscopy (AFM) and scanning electron microscopy (SEM), and we compared the inner capillary surfaces to rubbed glass; the capillaries had no discernible grooved structures as on the rubbed glass. Since SSY is known to exhibit natural planar anchoring on smooth glass surfaces \cite{nazarenko_surface_2010}, our observations of the capillary surface strongly suggest that degenerate planar boundary conditions are present on the inner surfaces of the cylinders and any anisotropic Rapini-Papoular type anchoring effect would be small \cite{ondris-crawford_curvature-induced_1993,crawford_saddle-splay_1995,NoteX}. We also considered alignment caused by flow during capillary filling. Loading capillaries with the LCLC in either the nematic or the isotropic phase resulted in the same type of director configurations. Further, since the filling flow is nearly perpendicular to the final alignment found at the capillary surface, flow alignment appears unlikely. Finally, we considered the possibility that a layer of molecules adsorbed to the capillary surface sets an easy access at the capillary surface during or shortly after filling. We exclude this possibility by cycling the filled capillary between nematic and isotropic phase and observing that the director at the capillary surface retains no memory from cycle to cycle \cite{NoteX}.

We measure the director angle, $\beta(r)$, directly by observing a flickering speckle pattern and its direction in the LC. The pattern originates from director field temporal fluctuations and accompanying fluctuations in the ordinary and extraordinary refractive indices which cause scattering  \cite{gennes_physics_1995}. These types of fluctuations of the director field have been exploited previously to measure the viscoelastic ratios of liquid crystals  \cite{hakemi_determination_1983,giavazzi_viscoelasticity_2014, zhou_elasticity_2014}. Our work follows~Ref.\cite{giavazzi_viscoelasticity_2014}, which proposed using videos of LC flickering to discern local orientation of the director field \cite{NoteX}. Flickering shape and direction depend on the local director field configuration and LC viscoelastic anisotropy.
  
\begin{figure}
 \includegraphics[width=1.0\linewidth]{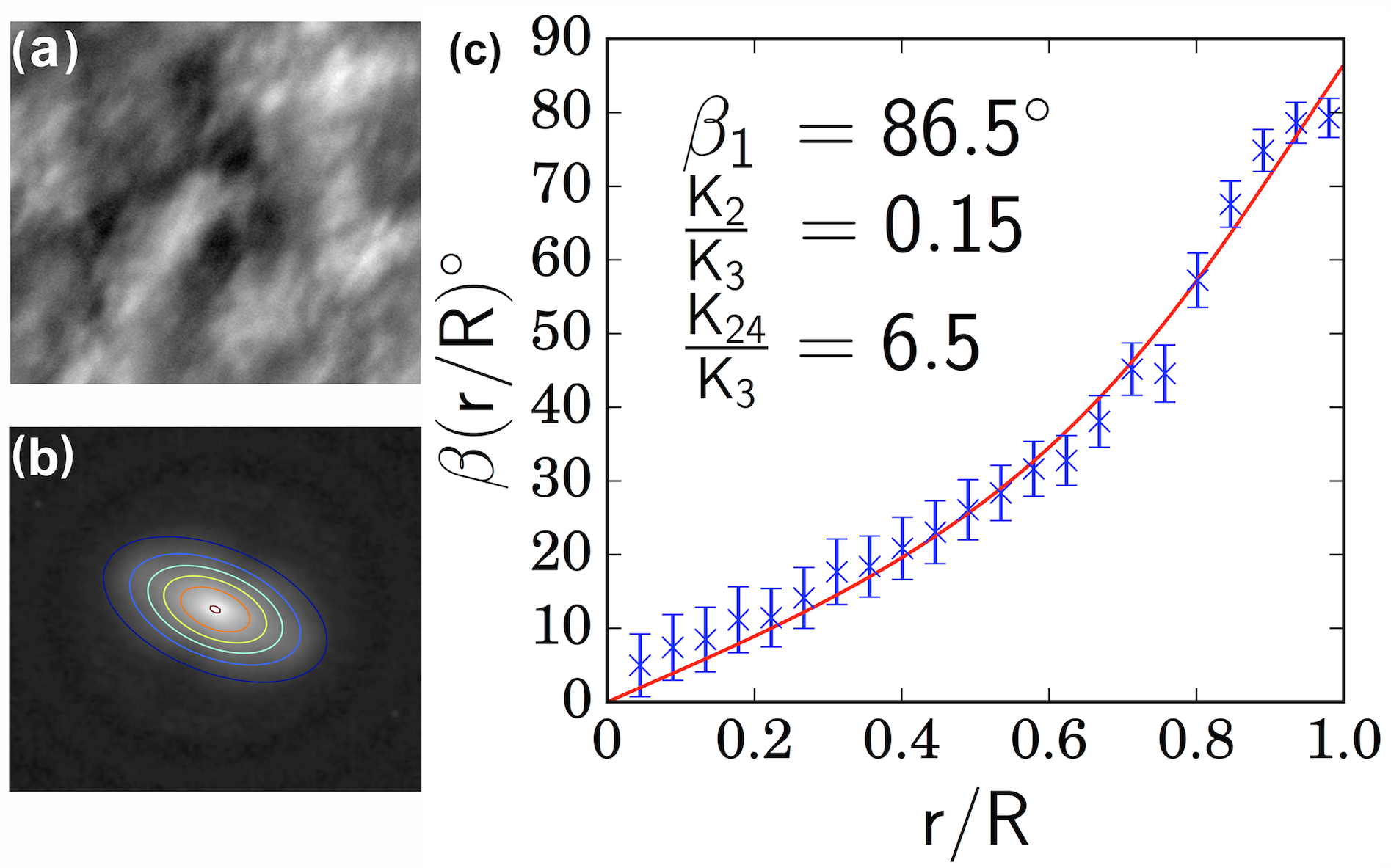}
 \caption{\label{Fig3} (Color online) Representative flickering measurements of 30\% wt./wt. SSY in a 90.6 \textmu m capillary at $25^{\circ}$C. {\bf (a)} Single frame from a movie cropped to a 20\textmu m square and after background subtraction (see main text). {\bf (b)} Averaged FFT of a movie containing many images of fluctuations and contour plot of a 2D Gaussian fit to the averaged FFT. The long axis of the fit is perpendicular to the dominant fluctuation direction and yields a measurement of $\beta$ for the image slice.  {\bf (c)} Fit of Eq.~(\ref{eqn:betas}) to $\beta(r)$ obtained by fluctuation measurements along the capillary radius. Error bars are the standard deviations in degrees of the angle found for the 2D gaussian fits as in {\bf (b)}. A nonlinear least-squares fit of the parameters $K_{2}/K_{3}$ and $K_{24}/K_{3}$ gives estimates of the elastic constant ratios ($6.5$ and $0.15$, respectively) for the sample. Across all measurements the average $K_{24}/K_{3} = 6.6$ and has a bounding interval $[3.8,9.4]$.}
 \end{figure}

The experimentally measured $\beta(r)$ for one of the five LCLC samples studied is shown in Fig.~\ref{Fig3}c. It is well fit by the calculated expression (Eq. 4), and the fitting provides experimental values for ratios of the twist-to-bend and saddle-splay-to-bend elastic constants. The twist-to-bend ratio is in close agreement with prior measurements   \cite{zhou_elasticity_2012}. Since $K_{24}/\sqrt{K_{3}K_{2}} \sim \tan(\beta_{1})$ when $K_{24}\gg K_{2}$, the fit values become increasingly sensitive to experimental uncertainties as $\beta_{1}\rightarrow \pi/2$. For example, the fit value of $K_{24}/K_3$ is sensitive to the uncertainty of the measured size of the capillary radius, $R$\cite{NoteX}. For the data in Fig.~\ref{Fig3}c, the capillary was measured to have a diameter of $90.6$~\textmu m to within $\approx \pm 0.4$~\textmu m. This relatively small uncertainty, however, leads to the large uncertainty we give for our estimate of $K_{24}/K_{3}$, i.e., $K_{24}/K_{3}$ has a mean value averaged across experiments of $6.6$ with bounding interval $[3.8,9.4]$.  By contrast, $K_{2}/K_{3}$ is a stiff parameter in the fit; it has a mean value averaged across experiments of 0.12 and a standard deviation $\sigma_{K_{2}/K_{3}}=0.04$ \cite{NoteX}. 
  
\begin{figure}[t]
 \includegraphics[width=1.0\linewidth]{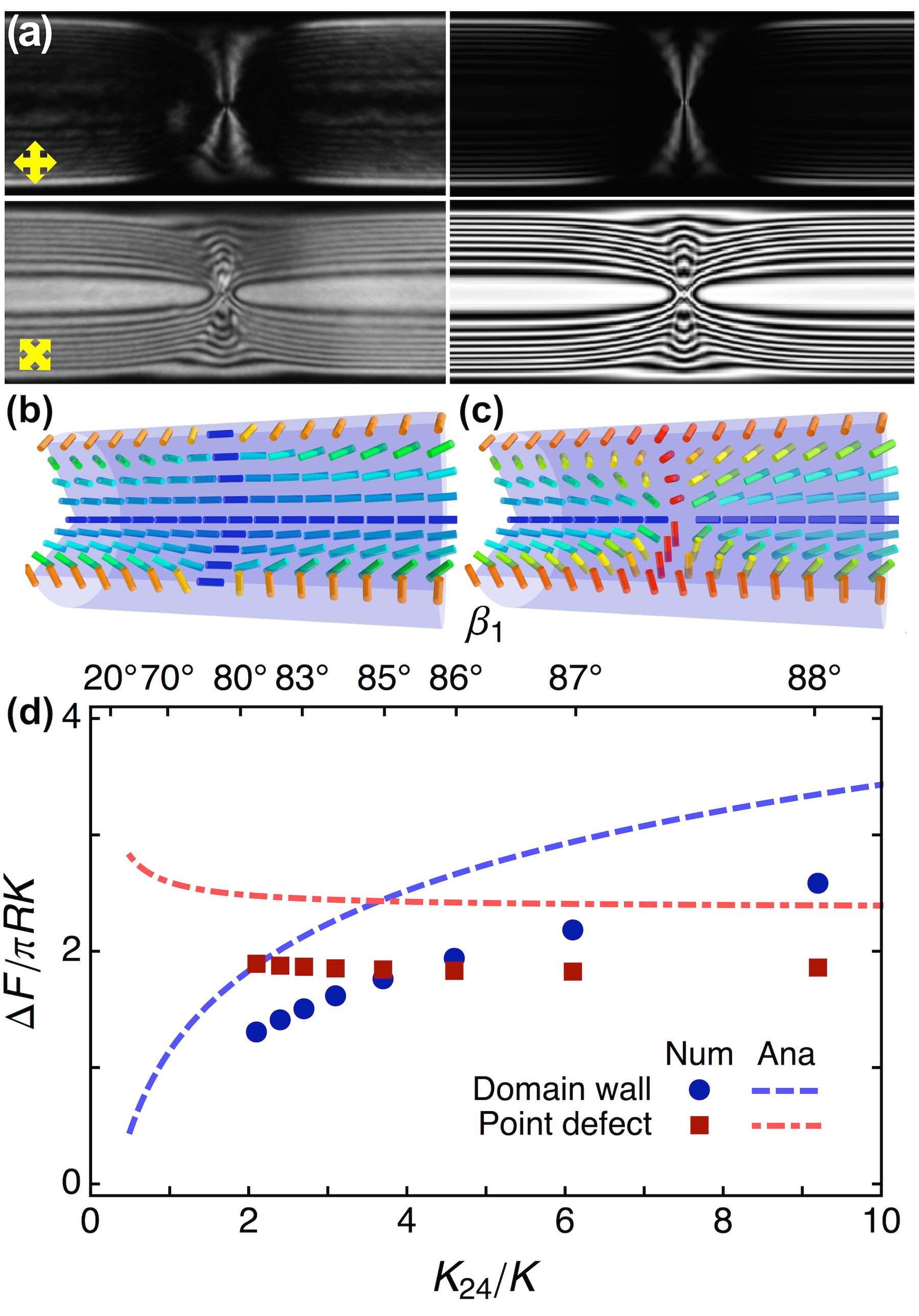}
 \caption{\label{Fig4} (Color online) {\bf (a)} Left, cross-polarized quasi-monochromatic optical images of singular point defects bordering ET regions of opposite handedness in 90 \textmu m capillary. Yellow arrows indicate the polarizer pass axis directions. Right, images reconstructed using Jones matrix calculations from numerically computed director fields of defects. {\bf (b)} A 3D cutaway view of a capillary with opposite-handedness ET regions separated by a wall defect. {\bf (c)} 3D cutaway view of a capillary with opposite-handedness ET regions separated by a point defect as imaged and simulated in {\bf(a)}. In both {\bf (b)} and {\bf (c)} the director field represents an LC with $K_{2}/K = 0.1$ and $K_{24}/K =4.6$, where $K_1 = K_3 \equiv K$ and the color scale is the same from Fig.~\ref{Fig1}b. {\bf (d)} Energies of the point and domain wall defects relative to the ET energy as a function of either $K_{24}/K$ or equivalently $\beta_1$, with $K_{2}/K = 0.1$. Points indicate numerical calculations and lines indicate analytical approxmations \cite{NoteX}; the latter have higher energy than the former but demonstrate similar qualitative behaviors.}
 \end{figure}

We also observed hedgehog defects associated with the ET configuration. In long capillaries, we typically observed ET domains of opposite handedness separated by chiral point defects. These defects were qualitatively proposed in Ref.\ \cite{melzer_optical_1977}. We observed annihilation of neighboring defects, indicating that they carry opposite topological charge\cite{NoteX}. The presence of nematic director singularities are apparent in Fig.~\ref{Fig4}a; bright-field microscopy \cite{NoteX} reveals dark spots from scattered light along the center of the capillary. Once found, we image the point defect under crossed-polarizers with the same illumination described above. We compare these experimental textures with those simulated numerically using Jones matrices \cite{NoteX}. The comparison requires a test director configuration, which we calculate using Eqs.~\ref{eqn:fgen} and~\ref{eqn:n}. For configurations in the presence of defects, however, the director depends on both $r$ and $z$; the boundary conditions at $z \rightarrow \pm\infty$ bring the director configuration back to ET configurations with opposite handedness. To arrive at an optimized guess, we solve the Euler-Lagrange equations numerically with a relaxational technique \cite{NoteX}. The configurations that emerge are very similar to what one gets if one takes the standard radial and hyperbolic hedgehogs and simply rotates all directors by $\pi /2$ about the $z$-axis \cite{NoteX}. This simple operation, which is guaranteed to preserve hedgehog charge, automatically produces opposite chirality on opposite sides along $z$ of the hedgehog defect regardless of the sign ($\pm 1$) of its charge. The topological charges of successive hedgehogs necessarily alternate in sign \cite{NoteX}.  Using $K_1=K_3\equiv K$, $K_2/K = 0.1$ and $K_{24}/K = 4.6$, numbers which are consistent with our measurements in the ET ground state, we observed remarkable agreement between experimental and theoretical textures (Fig.~\ref{Fig4}a).

In principle, smooth domain walls can also separate domains of opposite handedness, in which the escaped-twist configuration continuously untwists from one domain to the wall mid-plane and then continuously re-twists with opposite handedness into the other domain (see Fig. \ref{Fig4}b). In this case, throughout the mid-plane, the director would align along the capillary axis. However, in SSY, we have never experimentally observed such a domain wall structure. Defect energetics provide an explanation for this observation which has an interesting consequence. Again, we numerically calculate the configurations of both domain walls and point defects to obtain their energies \cite{NoteX}. For these calculations, we fix $K_2/K = 0.1$ in accordance with \cite{zhou_elasticity_2012} and our fluctuation experiments, and we allow $K_{24}$ to vary. As shown in Fig.~\ref{Fig4}d, point defects (domain walls) have lower energy than domain walls (point defects) for $K_{24}/K \gtrsim 4$ ($K_{24}/K \lesssim 4$). Using $K_3 = K = \SI{6.5}{\pico\newton}$ from   \cite{zhou_elasticity_2012} and $R = \SI{50}{\micro\meter}$, a typical dimensionless energy difference of $\Delta F/\pi RK = 0.1$ corresponds to $\Delta F = 2.5 \times 10^4 k_{B}T$, where $T = \SI{298}{\kelvin}$ is the experimental temperature. If $K_{24}$ is greater than the crossover value ${\approx}4K$, then, according to theory, one should not expect to observe smooth domain walls.  Thus, both our observations of defects (or lack thereof) and our energy analysis set $4$ as an approximate lower bound for $K_{24}/K$, in agreement with our fluctuation-measured value of $K_{24}/K = 6.6~[3.8,9.4]$.

In summary, we have completed an experimental and theoretical study of a lyotropic chromonic liquid crystal, Sunset Yellow, in its nematic phase and confined in a hollow cylinder with degenerate planar boundary conditions. The escaped-twist configurations found to form in the bulk require a large saddle-splay modulus, which we have measured. We also observed point defects in this system whose existence (compared to the absence of smooth domain walls) provides independent confirmation of $K_{24}$. In the future, it will be interesting to study and manipulate these chiral configurations and investigate their formation from the isotropic phase.

\begin{acknowledgments}
We thank Randall Kamien for helpful discussions and gratefully acknowledge financial support from National Science Foundation Grants DMR-1205463, DMR-1120901, DMR-1104707 and National Aeronautics and Space Administration Grant \#NNX08AO0G.
\end{acknowledgments}


%

\end{document}